\documentclass[twocolumn,10pt]{Paper}

\usepackage{epsfig} %% for loading postscript figures
\usepackage{float}
\usepackage{xcolor}
\usepackage{epstopdf}

\usepackage[labelfont=bf,textfont=md]{caption}
\usepackage[hidelinks]{hyperref}
\newcommand{\cc}[1]{\textcolor{black}{#1}}
\newcommand{\dd}[1]{\textcolor{black}{#1}}

\newenvironment{ddenvironment}{\par\color{black}}{\par}

\title{The Extreme Mechanics of Viscoelastic Metamaterials}

\author{David M.J. Dykstra, Shahram Janbaz and Corentin Coulais
    \affiliation{
	Institute of Physics\\
	University of Amsterdam\\
        Amsterdam, 1098 XH\\
        Netherlands\\
    }
}

\begin{document}

\maketitle

%%%%%%%%%%%%%%%%%%%%%%%%%%%%%%%%%%%%%%%%%%%%%%%%%%%%%%%%%%%%%%%%%%%%%%
%\dd{QUESTION: "strain rate" or "strain-rate"? I've changed everyting to "strain rate"}\\
\begin{abstract}
\textit{Mechanical metamaterials made of flexible building blocks can exhibit a plethora of extreme mechanical responses, such as negative elastic constants, shape-changes, programmability and memory. To date, dissipation has largely remained overlooked for such flexible metamaterials. As a matter of fact, extensive care has often been devoted in the constitutive materials’ choice to avoid strong dissipative effects. %Only recently has dissipation in mechanical metamaterials started to attract some attention: 
However, in an increasing number of scenarios, where metamaterials are loaded dynamically, dissipation can not be ignored. 
%\dd{Instead, it can be harnessed to obtain new functionalities.}
%on one hand demonstrating the shortcomings from ignoring dissipation in mechanical metamaterials 
%but on the other hand demonstrating the wide variety of additional functionalities which can be obtained by leveraging dissipation in mechanical metamaterials. 
In this review, we show that the interplay between mechanical instabilities and viscoelasticity can be crucial and can be harnessed to obtain new functionalities.
We first show that this interplay is key to understanding the dynamical behaviour of flexible dissipative metamaterials that use buckling and snapping as functional mechanisms. We further discuss the new opportunities that spatial patterning of viscoelastic properties offer for the design of mechanical metamaterials with properties that depend on loading rate.}
\end{abstract}

%%%%%%%%%%%%%%%%%%%%%%%%%%%%%%%%%%%%%%%%%%%%%%%%%%%%%%%%%%%%%%%%%%%%%%

%%%%%%%%%%%%%%%%%%%%%%%%%%%%%%%%%%%%%%%%%%%%%%%%%%%%%%%%%%%%%%%%%%%%%%
\section{Introduction}
Mechanical metamaterials exhibit exotic mechanical responses. 
Static responses of interest span a wide range of tunable behaviours, such as auxetic~\cite{lakes_foam,bertoldi_negative,babaee_3D}, programmable~\cite{florijn1,florijn2}, shape-changing~\cite{overvelde_reconf, coulais_multi}, non-reciprocal~\cite{coulais_recip} to chiral responses~\cite{frenzel_twist}, often by harnessing nonlinear mechanics and snap-through instabilities~\cite{florijn1,florijn2,rafsanjani_bistable,chi_multistable,restrepo_phase}. Interesting dynamical responses include shock absorption~\cite{frenzel_microlat,shan_trapping,correa_honeycombs,schaedler_metallic}, soliton propagation~\cite{deng_soliton,deng_gaps} and transition waves~\cite{Raney_propagation,nadkarni_unidir}.

\begin{figure}[b!]
\centerline{\psfig{figure=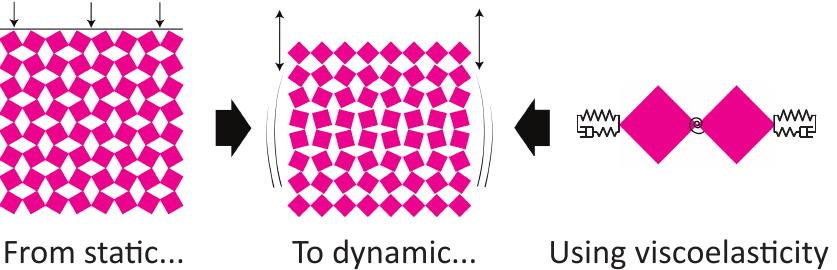,width=0.99\columnwidth}}
\caption{\textbf{Viscoelastic Metamaterials.} Viscoelasticity affects the dynamics of buckling and can host new responses and functionalities.}
\label{fig:Figure0}
\end{figure}

In many of these dynamical responses, dissipation plays a key role. Specifically, dissipation plays a key role in vibration transmission. As such, dissipative vibration transmissions has been studied extensively in acoustic metamaterials featuring wave propagation and band gap structures~\cite{merheb2008elastic,hussein2013damped,manimala2014microstructural,frazier2015viscous,wang2015wave,lewinska2017attenuation,ghachi2020optimization,miniaci2018design,banerjee2019waves,parnell_soft} as well vibration and shock control~\cite{hussein2013metadamping,bacquet2018metadamping,mu2020review,ji2021vibration,dalela2021review,al2022advances}. However, although the effect of the constitutive materials’ dissipation on mechanical instabilities has been well studied~\cite{kempner1954creep,nachbar1967dynamic,santer2010self,brinkmeyer2012pseudo,brinkmeyer2013pseudo,gomez_dynamics,urbach2020predicting,stein2021delayed}, the role of dissipation in the instabilities that underpin the extreme mechanics of mechanical metamaterials remains poorly understood (Fig \ref{fig:Figure0}).
%Although the effect of the constitutive materials’ dissipation on mechanical instabilities has been well studied~\cite{kempner1954creep,nachbar1967dynamic,santer2010self,brinkmeyer2012pseudo,brinkmeyer2013pseudo,gomez_dynamics,urbach2020predicting,stein2021delayed}, dissipation has remained overlooked for nonlinear metamaterials.
%As a matter of fact, extensive care has often been devoted in the constitutive materials’ choice to avoid strong dissipative effects.
%Only recently has shape-changing dissipation in mechanical metamaterials started to become understood. 
Yet, there is growing evidence that dissipation can play a key role in the response of mechanical metamaterials that are based on buckling~\cite{dykstra2019viscoelastic,che2019temperature}. In addition, a growing body of work has recently demonstrated that a wide variety of additional functionalities can be obtained by leveraging dissipation in mechanical metamaterials~\cite{stern2018shaping,che2019temperature,janbaz2020strain,bossart2021oligomodal}. 

In this short review, we discuss the interplay between mechanical instabilities and viscoelasticity and how this interplay can be harnessed to offer new functionalities that depend on loading rate.

%. We also discuss the new opportunities that spatial patterning of viscoelastic properties offer for the design of mechanical metamaterials with properties that depend on loading rate. 

\section{Delayed Snap-through Buckling by Dissipation}
Mechanical metamaterials often harness mechanical instabilities, including snap-through instabilities~\cite{bertoldi2017flexible}. In particular, multistable buckling paths can be harnessed to switch between \dd{different} responses~\cite{bertoldi2017flexible}. As such, we need to understand how dynamics and dissipation affect mechanical (snap-through) instabilities and multistability before we can develop new types of dissipative metamaterials.

%Mechanical metamaterials often harness mechanical instabilities~\cite{bertoldi2017flexible}. \sj{Here it is useful to mention why bisatbility is useful in shape changing metamaterials.} In particular, they often employ snap-through instabilities~\cite{bertoldi2017flexible}. As such, we need to understand how dynamics and dissipation affects mechanical (snap-through) instabilities before we can develop new types of dissipative metamaterials.  

On \dd{the} one hand, while quasi-static buckling analyses do not consider any timescale, even purely elastic snap-through instabilities show a very distinct time dependence, transitioning from a very slow state of storing elastic energy to a fast snap-through motion~\cite{gomez2017critical}. This \dd{snap-through motion} is what allows for instance the Venus flytrap  or the hummingbird to catch their prey~\cite{forterre2005venus,smith2011elastic}. \dd{Such snap-through motion can further be tuned by controlling the loading rate~\cite{liu2021delayed}.}

However, most solid materials are not purely elastic. Most solid materials are viscoelastic instead, \cc{and exhibit} both viscous and elastic characteristics when undergoing deformation. Viscous materials, like liquids, start flowing when a stress is applied. Purely elastic materials return to their original state once the stress is removed without dissipating any energy. \dd{Viscoelastic materials in turn} exhibit dissipation and a time-dependent stress-strain response. Specifically, viscoelastic materials stiffen with increasing strain rates. Basic models used for viscoelastic solids are discussed in the Materials \& Methods.

%Adding viscoelasticity complicates the dynamics of instabilities. Viscoelastic dissipation in instabilities has already been explored extensively since the pioneering works of Kempner in 1954 for beam buckling\cite{kempner1954creep} 

Adding dissipation complicates the dynamics of instabilities. Viscoelastic dissipation in instabilities has already been explored extensively since the pioneering works of Kempner in 1954 for beam buckling~\cite{kempner1954creep} and Nachbar and Huang in 1967 for snap-through instabilities~\cite{nachbar1967dynamic}. As one might expect, they showed that the buckling force increases with increasing loading rate, in line with what one might expect with linear viscoelastic materials.  

\begin{figure}[t!]
\centerline{\psfig{figure=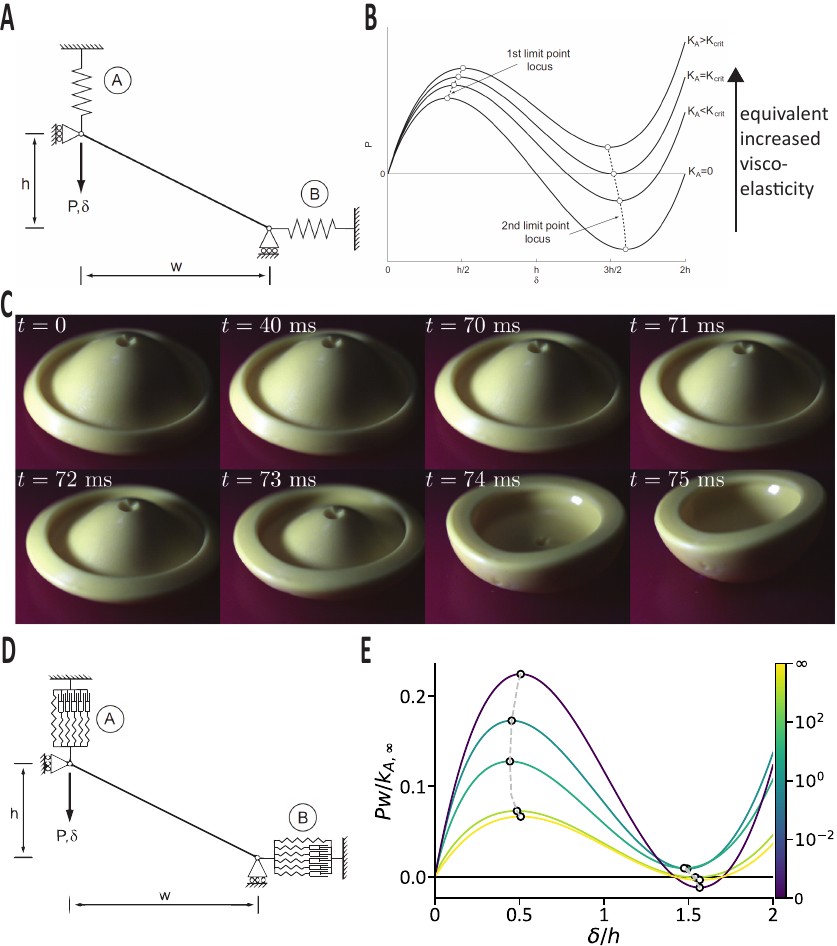,width=0.99\columnwidth}}
\caption{\textbf{Pseudo-bistability controlled by viscoelasticity.} %\textbf{Delayed buckling by dissipation.}
(A) A single rigid truss supported at both ends by elastic springs \dd{(Reproduced from~\cite{santer2010self} with permission from Elsevier)}. Changing the ratio of the spring stiffnesses can be considered equivalent to increasing the loading rate in a viscoelastic analysis. The equilibrium paths in (B) \dd{(Reprinted from~\cite{santer2010self} with permission from Elsevier)} show that both monostable and bistable configurations are possible. (C) A viscoelastic jumping popper shows pseudo-bistability  (Reproduced from~\cite{gomez_dynamics} under \href{https://creativecommons.org/licenses/by/4.0/}{CC BY 4.0}). (D) A viscoelastic version of the Von Mises Truss of (A), featuring Generalised Maxwell springs with 5 spring-dashpots \dd{(Modified from~\cite{santer2010self} with permission from Elsevier)}. (E) Force-Displacement curves due to moving at constant speed, with the colour indicating the total loading time in seconds. Both very high and very low speeds induce bistability, while intermediate speeds can generate monostability.}%(D) A laterally preconfined viscoelastic mechanical metamaterial for shock absorption experiences a snap-through instability only when compressed slowly (adapted from~\cite{dykstra2019viscoelastic}). 
\label{fig:Figure1}
\end{figure}

From a qualitative perspective, an equivalent quasi-static analysis can offer significant insights, such as the Von Mises Truss set-up explored by Santer in Fig. \ref{fig:Figure1}A~\cite{santer2010self}. Here, a rigid Von Mises truss is suspended by a vertical spring $A$ and a horizontal spring $B$ on the top and right, with spring stiffness $K_A$ and $K_B$ respectively. When pulling down the top of the Von Mises truss, it can be argued that increasing the ratio of $K_A/K_B$ is equivalent to increasing the strain rate in a viscoelastic analysis: spring A experiences a much larger deflection rate than spring B when the Von Mises truss is deflected downward, and since both springs are viscoelastic, the instantaneous spring constant or spring A is effectively larger than that of spring B. %Therefore, spring A might obtain a larger increase in effective stiffness due to viscoelastic effects. 
The relation between force $P$, deflection delta and $K_A/K_B$ is then captured in Fig. \ref{fig:Figure1}B~\cite{santer2010self}. Two major effects can be identified in Fig. \ref{fig:Figure1}B. First of all, as $K_A/K_B$ increases, not only does the buckling force of the first limit point increase, but so does its deflection. Second, as $K_A/K_B$ increases, the second limit points moves upward and stops crossing 0 above a critical $K_A>K_{crit}$.  At this point, the structure loses its bistability and becomes monostable.  In other words, a viscoelastic Von Mises Truss can be bistable if it is loaded slowly but monostable when it is loaded fast.%. However, if it is loaded fast, it becomes monostable instead and can snap back to its initial state.

%In other words, a viscoelastic structure can appear stable in its snapped state if it is quickly loaded. But after viscoelastic relaxation occurs, the snapped state will loose its stability and the structure will snap back into the initial state.

%In other words, a viscoelastic structure can appear stable in its snapped state if it is quickly loaded. But after viscoelastic relaxation occurs, the snapped state will loose its stability and the structure will snap back into the initial state.

%\sj{So, it it loses its bistability at high strain rates, how it can become useful for shape changing. Dose it prevents some local instabilities or specific events. Do you think we can discuss it here?}

%\dd{Question: Remake Fig 1. B by modelling viscoelastic snap-through ourselves?} 
 
The opposite effect can also occur. A viscoelastic structure can appear stable in its snapped state \dd{after it is loaded}. But after viscoelastic relaxation occurs, the snapped state will lose its stability and the structure will snap back into the initial state. This behaviour is in fact very common in viscoelastic snap-through instabilities and has been referred to as temporary bistability~\cite{santer2010self}, pseudo-bistability~\cite{brinkmeyer2012pseudo,brinkmeyer2013pseudo,gomez_dynamics}, acquired bistability~\cite{urbach2020predicting} or metastability~\cite{brinkmeyer2012pseudo}. For instance, viscoelastic jumping poppers (Fig. \ref{fig:Figure1}C) can appear bistable initially but may still snap back to the original state due to viscoelasticity~\cite{gomez_dynamics}.

\begin{figure*}[t!]
\centerline{\psfig{figure=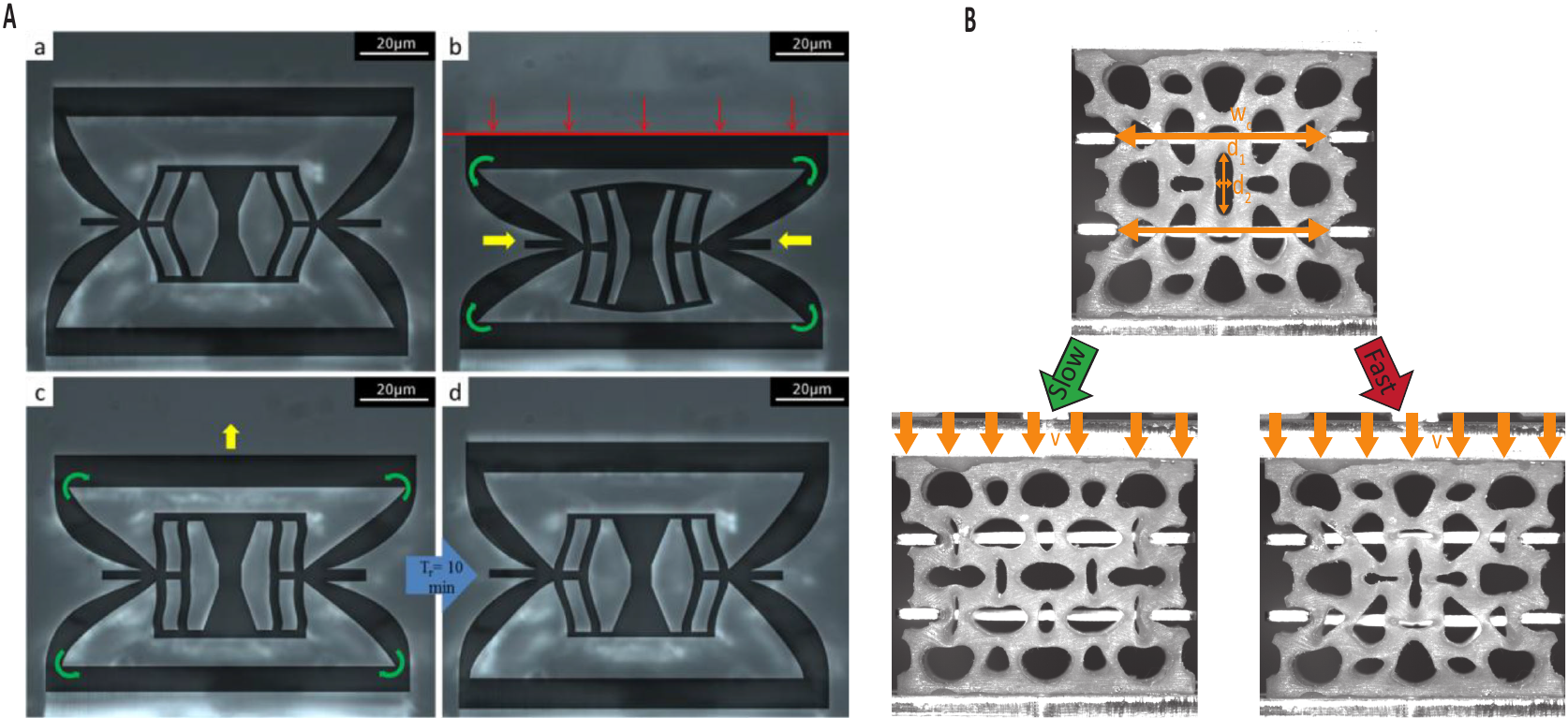,width=0.99\textwidth}}
\caption{\textbf{Delayed buckling by dissipation in mechanical metamaterials.} 
(A) A viscoelastic unit cell (a) after compression (b), showing initially bistable behaviour (c) before snapping back after relaxation (d) \dd{(Copyright Wiley-VCH GmbH. Reproduced from ~\cite{berwind2018hierarchical} with permission)}. (B) A laterally pre-confined viscoelastic mechanical metamaterial for shock absorption experiences a snap-through instability only when compressed slowly (Reproduced from~\cite{dykstra2019viscoelastic} with permission from ASME). }
\label{fig:Figure1-2}
\end{figure*}

However, while a quasi-static analysis can show some agreement with viscoelastic analyses, it can also differ substantially. %give us some indications of the behaviour of viscoelastic structures, a quasi-static analysis can also differ. 
To compare, we model the same Von Mises Truss of Fig. \ref{fig:Figure1}A but by \dd{fully} modelling the viscoelasticity of the springs (Fig. \ref{fig:Figure1}D) so that they faithfully describe the behaviour of a 3D printed rubber\dd{, which has also been explored in depth analytically by Gomez et al. \cite{gomez_dynamics}}. We do so with the Viscoelastic Solver in Abaqus 2021 \dd{(See Section\ref{sec:modelling})}, using a three element truss model (T2D2). Specifically, we replace springs A and B in Fig. \ref{fig:Figure1}A by viscoelastic springs using a viscoelastic 5-term Prony series, which is equivalent to a Generalised Maxwell-Model with 5 spring-dashpots~\cite{Christensenvisco} (see Materials \& Methods). For the strength and time scales of the Prony series, we assume that the springs are made from Stratasys Agilus30 material, a rubber-like material that is used in polyjet additive manufacturing and is commonly used as a prototypical example of a highly viscoelastic solid~\cite{dykstra2019viscoelastic,janbaz2019ultra,mehta2019fabrication,bossart2021oligomodal,yu2021exploration,lu20213d}. The strength and time scales of the Prony series have been adapted from~\cite{dykstra2019viscoelastic} and can be found in the Materials \& Methods. We employ $w/h=10$ and $K_A/K_B=0.05$. We then move the truss down with a constant rate and track the force-displacement response at various loading rates in Fig. \ref{fig:Figure1-2}E. 

When going from small loading rates in yellow to intermediate loading rates in green and cyan, we observe a change from bistability (\dd{force goes through zero three times}) to monostability (\dd{force crosses zero only once}). %This is in fact opposite to pseudo-bistability, in that a structure which is bistable in its relaxed condition becomes monostable due to viscoelasticity. This shows that it is not self-evident to predict in what way viscoelasticity affects the stability of a buckling structure. 
Also, the reaction force $P$ increases over the entire loading regime when increasing the loading rate. Both of these phenomena were also captured by the quasi-static analysis in  Fig. \ref{fig:Figure1}B. However, when we further increase the loading rate, we again identify bistability, which is not captured by a simple quasi-static analysis. Relaxation from this state could induce another state of intermediate monostability instead, so the same structure could also capture pseudo-bistability. This shows that it is not self-evident to predict in what way viscoelasticity affects the \dd{pseudo-}stability of a buckling structure. %\dd{In fact, multiplying the force curve of the quasi-static loading case (yellow) with the ratio between the instantaneous and relaxed spring stiffness, results in the instantaneous loading force curve (dark blue). }%In fact, when you multiply the force curve of the quasi-static loading case (yellow) with the ratio between the instantaneous and relaxed spring stiffness, you end up with the instantaneous loading force curve (dark blue). %the forces in the instantaneous case (dark blue) are a direct multiplication of the quasi-static loading case (yellow). This is because there is no change in apparent stiffness in either case. 
Moreover, while the quasi-static analysis in Fig. \ref{fig:Figure1}B suggests that the first limit point will shift to the right when increasing the equivalent loading rate, the viscoelastic analysis in  Fig. \ref{fig:Figure1}E shows the opposite effect. This shows that it is not trivial to identify a priori whether viscoelasticity will increase or decrease the buckling deflection either.
%However, while it is very logical that viscoelasticity can increase the buckling force% and deflection
In both of these cases, increased viscoelasticity, as identified by a higher loading rate, has led to an increase in buckling force. 

Yet, the opposite can also be true.  Stein-Moldavo et al. recently showed that the buckling pressure of spherical shells could also drop due to viscoelastic effects~\cite{stein2021delayed}. When a negative pressure was kept stable just below the elastic buckling limit, a viscoelastic shell would still buckle after relaxation. This is because the relaxation induces additional geometric imperfections, which in turn leads to a reduction in the buckling pressure. However, depending on the direction of the imperfections, the result could also act opposing and make the system more bistable~\cite{liu2021effect}. 
%In summary, there is a well established understanding of the role of viscoelasticity on instabilities of flexible structures. 

In summary, the interplay between instabilities, imperfections and viscoelasticity all contribute to shaping how snap-through buckling is affected by viscoelasticity.

%In comparison, comparably little is known on metamaterials, which consist of many such flexible building blocks. 
In turn, this knowledge of viscoelasticity-induced pseudo-stability can be used to understand how 
%It turns out that the viscoelasticity-induced pseudo-stability can in fact greatly affect 
the behaviour of mechanical metamaterials is affected by viscoelasticity. First of all, Li et al. have shown that viscoelastic dissipation can delay buckling in viscoelastic metamaterials, which in turn can delay a shift from non-auxetic to auxetic in mechanical metamaterials~\cite{li20213d}. Moreover, Berwind et al. showed in Fig. \ref{fig:Figure1-2}A that pseudo-stability can occur in mechanical metamaterials just as it could for simpler visco-elastic snap-through instabilities~\cite{berwind2018hierarchical}. \dd{Pseudo-bistability can even be harnessed to program a morphing pattern of viscoelastic snap-through shells~\cite{chen2022spatiotemporally}}. On the other hand, Dykstra et al. showed in Fig. \ref{fig:Figure1-2}B that viscoelasticity can prevent a mechanical metamaterial featuring snap-through instabilities from snapping altogether under large strain rates~\cite{dykstra2019viscoelastic}. From a shock damping perspective, this means that a higher strain rate could actually induce poorer shock damping performance when no snap-through occurred. The strain at which this transition occurred was then tailored by shape-changing. This showed the very careful interrelation between geometry and dissipation in viscoelastic metamaterials. 

%A delay in buckling %This study has focussed on a metamaterial with a small number of unit cells. The role of the interplay between viscoelastic 

In summary, delayed snap-through and modification of stability by viscoelasticity can occur because different parts of the structure or the metamaterial can experience different strain rates. As a result, these different parts experience different levels of stiffening induced by viscoelasticity and this modifies the energy landscape. In the second section, we will show that suitable spatial patterning of the viscoelastic properties allows to further control how the energy landscape depends on strain rate.  

\section{Dissipation to Seed Dynamic Imperfections}
\begin{figure*}[t!]
\centerline{\psfig{figure=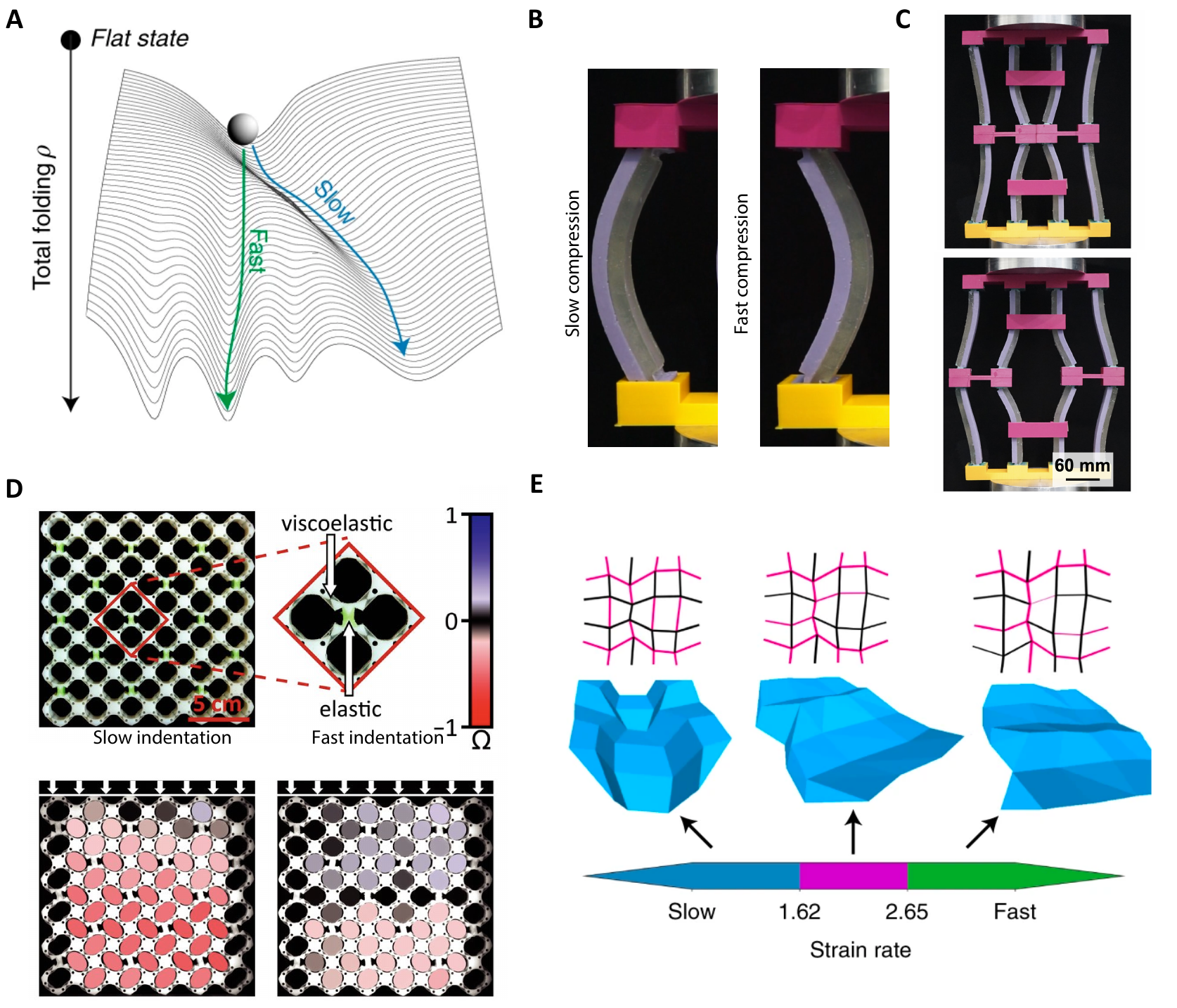,width=0.99\textwidth}}
\caption{\textbf{Dissipation to seed dynamic imperfections.} (A) Viscoelasticity and loading rate can be used to navigate an energy landscape with multiple valleys \dd{(Reproduced from from~\cite{stern2018shaping} under \href{https://creativecommons.org/licenses/by/4.0/}{CC BY 4.0})}. (B) Elastic-viscoelastic bi-beams can switch buckling direction when changing loading rate \dd{(Reproduced from~\cite{janbaz2020strain} under \href{https://creativecommons.org/licenses/by-nc/4.0/}{CC BY-NC 4.0})} and (C) can allow switching between auxetic and non-auxetic when turned into a mechanical metamaterial \dd{(Reproduced from~\cite{janbaz2020strain} under \href{https://creativecommons.org/licenses/by-nc/4.0/}{CC BY-NC 4.0})}. (D) Loading rate can also be used to switch modes altogether in metamaterials with a distinct number of global modes and both elastic and viscoelastic hinges \dd{(Reproduced from~\cite{bossart2021oligomodal} under \href{https://creativecommons.org/licenses/by-nc-nd/4.0/}{CC BY-NC-ND 4.0})}. (E) An origami sheet featuring a higher number of modes and both elastic and viscoelastic hinges can also experiences alternative modes at intermediate loading rates \dd{(Reproduced from from~\cite{stern2018shaping} under \href{https://creativecommons.org/licenses/by/4.0/}{CC BY 4.0})}.}
\label{fig:Figure2}
\end{figure*}

Flexible metamaterials based on buckling and snap-through instabilities exhibit complex energy potentials---with mountains and valleys. We argue in this section that viscoelastic dissipation can be used to introduce a new stable bifurcation state~\cite{alhadidi2021new}. This in turn is an efficient way to control how metamaterials navigate such landscapes, based on the loading rate  (Fig. \ref{fig:Figure2}A)~\cite{stern2018shaping}. As such, the deformation pathways and ensuing shape-changes and effective properties of the metamaterials can depend dramatically on whether the metamaterials are loaded fast or slow. 

The control over such deformation pathways can be achieved by spatially patterning the viscoelasticity of the flexible building blocks. The simplest example is perhaps that of Euler beam buckling. In such a case, the patterning of the viscoelastic relaxation strength allows to induce an imperfection that depends on strain rate~\cite{janbaz2020strain} (Fig. \ref{fig:Figure2}B). In practice, take a column split in two parts: the left (right) half of the column is made of a material with a small (large) viscoelastic relaxation strength. When the column is loaded along its axis, it will buckle to the left (right) when loaded slower (faster) than the typical relaxation timescale of the more viscoelastic material. This happens due to a shift in the neutral axis. When loaded slowly (fast), the viscoelastic material is much softer (stiffer), so the neutral axis shifts left (right) of the point of load application and the beam therefore buckles to the left (right). \dd{When sufficiently compressed, nonlinear deformation prevents snap-back, implying a permanently stable bifurcation.} This idea can be applied to construct flexible structures that exhibit switching between auxetic and non-auxetic behaviour (Fig. \ref{fig:Figure2}C) or show an apparent negative viscoelasticity---where the apparent stiffness is lower for larger loading rate and larger for lower loading rates~\cite{janbaz2020strain}.
%by preventing the post buckling of beam elements as a result of a contact design
% \dd{EXPLAIN WHAT NEGATIVE VISCOELASTICITY MEANS}.

The section above shows that the direction of buckling of a beam or a hinge can be controlled by spatially patterning its viscoelastic properties. In this case there is a single deformation mode (the out-of-plane bending) and one controls its direction by inducing a strain-rate dependent imperfection. %\sj{It is better to star this paragraph from here}. 

In the following, we also discuss a different approach, which consists of using structures with more than one deformation mode and where the spatial patterning allows to selectively actuate the mode of choice based on loading rate. Architectures that have been used for such an approach are oligomodal metamaterials (Fig. \ref{fig:Figure2}D)~\cite{bossart2021oligomodal} or origami with multiple folding branches (Fig. \ref{fig:Figure2}E)~\cite{stern2018shaping}. The idea is the following: a metamaterial has a distinct number of global deformation modes (or folding branches for origami), such as the metamaterial visualised in Fig. \ref{fig:Figure2}D which has two modes. One mode only actuates viscoelastic hinges, all of which act soft during slow loading of the structure. The other mode also actuates elastic hinges, which are much softer than the viscoelastic hinges during fast loading of the structure. As a result, fast loading will energetically favour deforming the mode featuring elastic components. Where there are more modes available, it is then also possible to obtain additional deformation modes at intermediate loading rates (Fig. \ref{fig:Figure2}E)~\cite{stern2018shaping}. %When using the languages of energy landscape, 

However, %while switching the direction of a mode (Fig. \ref{fig:Figure2}B)\cite{janbaz2020strain} has no energetic preference in the idealised case, changing the nature of the mode often does (Fig. \ref{fig:Figure2}DE)\cite{bossart2021oligomodal,stern2018shaping}. 
using only viscoelasticity is often not enough to navigate the energy landscape effectively and \cc{to} allow for a switchable actuation of the mode of choice~\cite{bossart2021oligomodal}. In some cases, some modes tend to have a lower energy barrier and as such can be preferentially actuated regardless of the rate of actuation or of the mismatch of viscoelasticity. Imperfections---traditionally considered a hazard in most mechanical systems---tailored to the preferred mode, can be used to tweak the energy landscape and hence to alleviate this problem. 
In quasi-static metamaterial designs they have allowed branching into specific chosen bifurcation paths~\cite{janbaz2019ultra,gavazzoni2022cyclic}, while they can enable switching between modes when combined with viscoelasticity~\cite{bossart2021oligomodal}. This means that imperfections can also be programmed specifically to change the loading rate at which mode switching occurs.

In summary, we have discussed how spatial patterning of viscoelastic relaxation strength allows to dynamically tune buckling imperfections or relative mode stiffness. We have discussed a few examples where this approach has been used to create metamaterials with shape-changes that depend on loading rate. \dd{In the next section, we will show that stored energy combined with viscoelasticity can also induce extreme mechanical properties.}

\begin{ddenvironment}
\section{Extreme Mechanics due to Negative Stiffness}

Traditionally, % materials provide a compromise between its constituents. 
the mechanical properties of \cc{a} composite material typically cannot surpass those of its constituents~\cite{lakes2002dramatically,kochmann2017exploiting}. For example, for the Young's modulus, upper and lower bounds for laminates are provided by the Voigt~\cite{voigt1889ueber} and Reuss~\cite{reuss1929berechnung} bounds, while upper and lower bounds for isotropic composite materials are provided by the Hashin-Shtrikman bounds~\cite{hashin1963variational} (Fig. \ref{fig:FigureNeg}A). A similar compromise is present for damping. As a result, a combination of high damping and high stiffness seems to be mutually exclusive~\cite{kochmann2017exploiting}. 

%The two materials that make up the  These bounds suggest that materials that combine high stiffness and high damping cannot be created. 

However, these bounds all assume that the constituent materials have positive definite elastic moduli, also known as positive stiffness~\cite{lakes2002dramatically,jaglinski2007composite,kochmann2014stable,fritzen2014material,kochmann2017exploiting}. In other words, they assume that the constituents do no not contain stored energy. However, while a freestanding material made out of a single material must display positive stiffness, confined systems can store energy. For example, post-buckled systems can display negative stiffness by releasing mechanical energy~\cite{kochmann2017exploiting}. Viscoelasticity in turn stabilises this state of negative stiffness~\cite{wang2004negative,wang2004stable,drugan2007elastic,jaglinski2007composite,kochmann2014rigorous,kochmann2014stable,kochmann2017exploiting}. In turn, the negative stiffness in these post-buckled systems can be harnessed to generate extreme damping~\cite{dong2012advanced,kalathur2013column,kochmann2017exploiting}. 

Furthermore, confined material unit cells can also display negative stiffness. In a notable example, Jaglinski et al. showed in Fig. \ref{fig:FigureNeg}B that a BaTiO$_3$ unit cell could transition from positive to negative stiffness when it is cooled without allowing deformation~\cite{jaglinski2007composite}. When the normalised temperature $T_N$ becomes negative, the Landau energy~\cite{landau1936theory} as function of strain $\epsilon$ becomes non-positive definite, implying negative stiffness (Fig. \ref{fig:FigureNeg}B). While such a negative stiffness unit cell would be unstable in an unconfined configuration, Jaglinski et al. embedded such unit cells in a tin matrix, creating a mechanical metamaterial featuring negative stiffness inclusions~\cite{jaglinski2007composite}. As a result, they were able to generate extreme damping in cyclic loading (Fig. \ref{fig:FigureNeg}C)~\cite{jaglinski2007composite}. In turn, it even allowed them to generate a viscoelastic modulus greater than diamond, the hardest natural material known to \cc{humankind} (Fig. \ref{fig:FigureNeg}C)~\cite{jaglinski2007composite}. 

Other authors have shown similar results, both theoretically and experimentally: negative stiffness stabilised by viscoelasticity can generate extreme mechanical properties, including extreme mechanical damping, extreme viscoelastic modulus and extreme thermal expansion~\cite{wang2004negative,wang2004stable,drugan2007elastic,jaglinski2007composite,kochmann2014rigorous,kochmann2014stable,kochmann2017exploiting}. This shows the great potential of using negative stiffness in viscoelastic metamaterials.

\begin{figure*}[t!]
\centerline{\psfig{figure=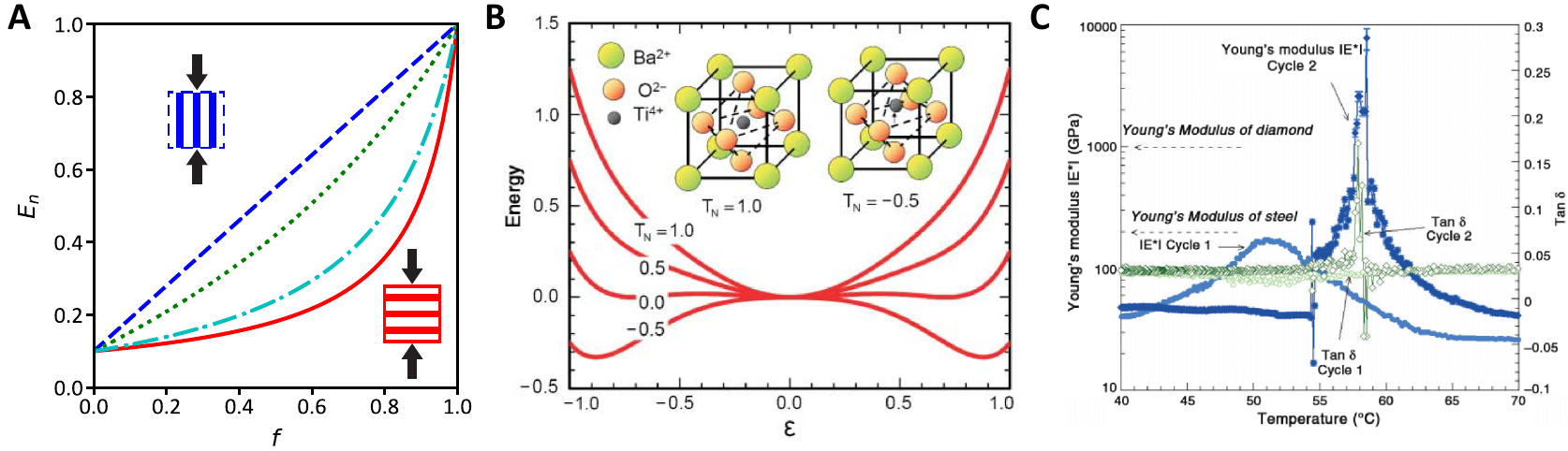,width=0.99\textwidth}}
\caption{\dd{\textbf{Negative Stiffness Inducing Extreme Dynamic Properties.} (A) Young's modulus bounds for composite materials made out of constituents with positive definite elasticity. The Young's modulus for laminates is bound between the Voigt~\cite{voigt1889ueber} (blue dashed line and inset) and Reuss~\cite{reuss1929berechnung} (red solid line and inset) bounds. The Young's modulus for isotropic materials is bound between the upper (green dotted line) and lower (cyan dash-dotted line) Hashin-Shtrikman~\cite{hashin1963variational} bounds. The two constituents have a stiffness ratio of 10 and Poisson's ratio $\nu=0.3$. $E_n$ is the composite Young's modulus normalised with the Young's modulus of the stiff constituent. $f$ is the volume fraction of the stiff constituent. (B) Landau energy function of strain $\epsilon$ and normalised
temperature $T_N = (\alpha \gamma/\beta^2)(T – T_1) –
0.25$, with unit cells of BaTiO$_3$ in
cubic and tetragonal phases. $\alpha$, $\gamma$,
and $\beta$ are constants that depend
on the material (Reproduced from \cite{jaglinski2007composite} with permission from AAAS). (C) Young's modulus $|E^*|$ and viscoelastic
damping $tan \delta$ of tin-BaTiO$_3$ composite metamaterial, showing extremely high modulus over a range of temperatures. $\delta$ is the phase angle between stress and strain. Young's moduli of the constituents are BaTiO$_3$, 100 GPa; and tin, 50 GPa (Reproduced from \cite{jaglinski2007composite} with permission from AAAS).}}
\label{fig:FigureNeg}
\end{figure*}
\end{ddenvironment}

\section{Modelling Viscoelastic Metamaterials}
\label{sec:modelling}
Another issue that comes with designing nonlinear viscoelastic metamaterials, is that of computational effort. While modelling quasi-static or linear systems require relatively little computational effort, modelling nonlinear dissipative systems often requires computationally expensive nonlinear dynamics. In most cases, the response can fully and accurately be modelled using nonlinear dynamic finite element methods. For relatively smooth dynamic analysis, implicit finite element methods can be used. \dd{Implicit dynamic analyses are analyses where relatively large time steps can be taken which each have to converge individually within numerical accuracy to the exact solution}. We have provided the codes for two of such cases on Zenodo~\cite{Zenodo2022}, corresponding to the analyses of Fig. \ref{fig:Figure1}DE and Fig. \ref{fig:Figure2}BC, which have been solved using the viscoelastic solver in Abaqus (Simulia). For more complicated dynamic analyses, explicit analyses may be required\dd{. Explicit dynamic analyses are analyses, which use very small time steps that do not need to converge to the exact solution individually. Instead, such analyses may jump around the exact solution from one step to another.} While implicit dynamic analyses can already be computationally expensive for a detailed mesh, explicit dynamics can be orders of magnitude more expensive when modelling long \dd{model time} durations. Often, using such dynamic finite element solvers is not computationally viable for more complicated systems, parametric studies or optimisation. As such, reducing this computational effort is another key point to advance the field of viscoelastic shape-changing metamaterials. 

%Another issue that comes with designing nonlinear viscoelastic metamaterials, is that of computational effort. While modelling quasi-static or linear systems require relatively little computational effort, modelling nonlinear dissipative systems often requires computationally expensive nonlinear explicit dynamics (\sj{We can also use Abaqus Standard Viso solver while we can neglect the effect of mass}). In some cases, the response can fully and accurately be modelled using nonlinear explicit \sj{I used standard/visco} finite element methods~\cite{janbaz2020strain}, but this is often not computationally viable for more complicated systems. 

Several strategies have already been identified. For one, many dissipative shape changes \dd{without inertial effects} can be approximated with \dd{fully elastic quasi-static analyses, by scaling local time-dependent material stiffnesses with a fixed value. Such analyses can} qualitatively match the key shape-change, which has been done among other for the results in Fig. \ref{fig:Figure1}AB~\cite{santer2010self} and Fig. \ref{fig:Figure2}D~\cite{bossart2021oligomodal}. However, these analyses only give a qualitative description and can fall short in obtaining the requested response~\cite{bossart2021oligomodal}. On the other hand, a simplified model employing full nonlinear dynamics, such as that of Fig. \ref{fig:Figure1}DE, can also offer qualitative agreement but may still fall short quantitatively~\cite{dykstra2019viscoelastic}. In such cases, the challenge remains in finding a minimum viable model without quantitative sacrifice. For systems that are overdamped or quasi-static, a viscoelastic finite element solver which neglects the effects of mass can prove viable (Fig. \ref{fig:ss}~\cite{janbaz2020strain}). Such solvers are available in both an implicit and explicit variant. For these overdamped or quasi-static analyses, neglecting the effects of mass can lead to multiple orders of magnitude improvement in computational effort, without quantitative sacrifice. In cases where mass is not negligible or where many unit cells have to be analysed simultaneously, another option would be to model mechanical metamaterials as a continuum. Statically, a conformal mapping approach has showed excellent quantitative agreement in modelling mechanism based metamaterials~\cite{czajkowski2022conformal}. More-over, Glaesener et al. have even shown that viscoelastic truss-based metamaterials can be modelled as time-dependent continua~\cite{glaesener2021viscoelastic}. In doing so, they managed to capture the full nonlinear dynamics of the metamaterials accurately. 

These works have shown that various strategies can be taken to model viscoelastic metamaterials more efficiently.  These strategies, and more efficient efficient strategies still, will allow us to predict and design a new generation of viscoelastic metamaterials. 

\section{Outlook}
%The use of dissipation in mechanical metamaterials is starting to show its potential with viscoelasticity but the possibilities are endless. 
We have reviewed here how dissipative effects can be used to control the extreme mechanics of metamaterials in the context of viscoelasticity. Further progress could be achieved in this direction by further improvements in additive manufacturing~\cite{skylar2019voxelated}, material science~\cite{hur2011three} and computational design methods. In particular, better control of multimaterial printing of materials with strong contrast in their viscoelastic properties could 
help in taking the ideas presented in this article further---one could imagine more complex geometries \dd{or different ways to embed internal energy}. Printing at the microscale could help in designing a wider range of shape-changing metamaterials with multiple functionalities. \dd{One could also harness viscosity to turn fluids subjected to acoustic vibrations into mechanical metamaterials~\cite{sehgal2022viscosity}}. 
%our control of spatial patterning of viscoelastic properties are fueling exciting new %developments. And these possibilities are not limited to viscoelasticity which has been the main focus of this review so far. 

One can also look beyond the case of viscoelasticity treated here to manage the energy landscape of adaptive and interactive materials and materials systems~\cite{walther2020responsive}. For instance, Che et al.~\cite{che2019temperature} showed that temperature can be used to expand the design space of the previously discussed pseudo-bistability in mechanical metamaterials~\cite{brinkmeyer2012pseudo,brinkmeyer2013pseudo,gomez_dynamics}. In particular, they showed that by changing the temperature, they could control the amount of time it took to snap-back. This is because the viscoelastic properties of rubbers are highly temperature dependent. Further down the line, one could envision that recent ideas on odd viscoelasticity~\cite{scheibner2020odd,brandenbourger2021limit,fodor2021optimal,banerjee2021active,lier2021passive,tan2021development}---where the viscoelastic tensors are not symmetrical and which have so far required active elements---could be realised via suitable combinations of purely passive viscoelastic elements.

Alternatively, dissipation can be harnessed from different mechanisms than viscoelasticity. For instance, internal adhesive contacts can be used to generate extreme dissipation and recovery after loading of nanofoams~\cite{park2020strong,white2021interpenetrating}. Active dissipation on the other hand can be used to control waves~\cite{brandenbourger2019non,scheibner2020odd} or achieve robotic functionalities~\cite{brandenbourger2021limit} in mechanical metamaterials. Otherwise, when combining elastomers with Low Melting Point Alloys, plasticity and heat be used to generate reversible plastic deformation~\cite{hwang2022shape}. This generates a response similar to what can be obtained using shape memory materials. One can also think of a plethora of alternative ways to harness dissipation, including for instance piezoelectric~\cite{shi2019architected}, magnetic~\cite{chen2021reprogrammable,liang2022phase}, photo-responsive~\cite{patel2022photo} or superelastic~\cite{yang2022towards} materials or fluidic devices~\cite{vasios2020harnessing}. \dd{In an even broader sense, nearly any type of time-dependent mechanical properties can be exploited for 4D printing: 3D printing with a fourth dimension in time~\cite{momeni2017review,yang20194d}. 4D printing can be used among others to generate morphing or reconfigurable structures, where viscoelastic creep and stress relaxation can play a vital role~\cite{yang20194d}.}

With so many possibilities to harness dissipation in the design of mechanical metamaterials, the next step is to go from generating new types of mechanical behaviour to new functionalities to actual applications. \dd{Obvious applications include using extreme vibration and shock damping due to negative stiffness in e.g. aerospace, automotive or high-tech machinery~\cite{kochmann2017exploiting}. However, viscoelastic metamaterials also show potential in} synthetic soft tissues and implants~\cite{saidy2019biologically}. Dissipative materials can also be beneficial to enrich the design of machine-like devices. For example, the locomotion of soft robots can be influenced by embedding dissipative components in their bodies~\cite{brandenbourger2021limit}. For instance, the control of motion can be reduced to the control of flow rate in pneumatic soft actuators~\cite{roche2014bioinspired,vasios2020harnessing}. For example, at larger scales, dissipative mechanisms in wearable devices and soft exo-suits~\cite{sanchez2021textile} can promote the design of such soft systems by reducing the duty of active elements and annexing memory.
%Besides these examples, the potential of using dissipative metamaterials will be endless along with the dramatic engineering advancements.

%and soft robotics\cite{vasios2020harnessing} offer exciting potential use cases.\dd{SHOCK AND VIBRATION DAMPING HAVE SPECIFICALLY ALREADY BEEN DONE}%, but dissipative metamaterials have the potential for many different applications such as soft robotics~\cite{}.  

%\section{Viscoelastic papers to include}

%Visco-thermal effects in acoustic metamaterials: from total transmission to total reflection and high absorption~\cite{moleron2016visco}. \\

%Force reversal and energy dissipation in composite tubes through nonlinear viscoelasticity of component materials~\cite{sedal2020force}\\

\
\appendix
\begin{center}
\section*{MATERIALS \& METHODS}
\end{center}
\bigbreak

\section*{Viscoelastic Material Models}
\begin{figure}[h!]
\centerline{\psfig{figure=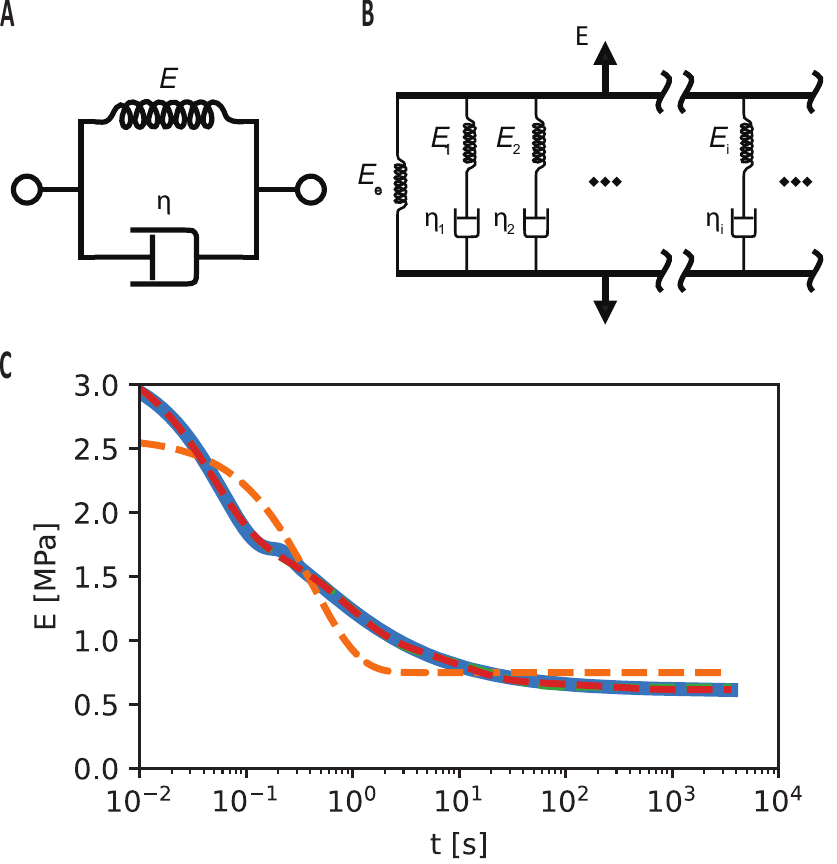,width=.99\columnwidth}}
\caption{\textbf{Viscoelastic material models.}(A) Representation of the Kelvin-Voigt model of viscoelasticity. (B) Representation of the Generalised Maxwell model. (C) Relaxation test for a dogbone sample of Agilus 30 3D printed material. Time-dependent Young's modulus vs. relaxation time. The thick blue line indicates the test results, the thin dashed orange, green and red curves present fitted data with $N=1$, $N=3$ and $N=5$ respectively. }
\label{fig:SI1}
\end{figure}

The two most commonly used viscoelastic material models for viscoelastic solids are the Kelvin-Voigt model (Fig. \ref{fig:SI1}A) and the Generalised Maxwell model (Fig. \ref{fig:SI1}B)~\cite{lakesvisco}.

\begin{table*}[h!]
\caption{Fitted viscoelastic material properties for Agilus 30, with $N$ = 1, $N$ = 3 and $N$ = 5}
\begin{center}
\label{tab:relax}
\begin{tabular}{|c c c| c c c c c | c c c c c |}
\hline
$N$ & $E_0$ & $E_e$ & $\beta_1$ & $\beta_2$ & $\beta_3$ & $\beta_4$& $\beta_4$ & $t_1$ & $t_2$ & $t_3$  & $t_4$ & $t_5$ \\
& [MPa] & [MPa] & & & & & & [s] &[s] &[s] &[s] &[s]\\
\hline
1 & 2.58 & 0.75 & 0.71 & & & & & 0.43 & & & & \\
3 & 3.25 & 0.64 & 0.45 & 0.26 & 0.10 & & & 0.047 & 0.97 & 18 & & \\
5 & 3.27 & 0.62        & 0.43 & 0.11 & 0.05 & 0.047 & 0.015         & 0.04 & 3.94 & 27.1 &  0.65 & 323\\
\hline
\end{tabular}
\end{center}
\end{table*}

The Kelvin-Voigt model (Fig. \ref{fig:SI1}A) consists of a spring, with a dashpot in parallel and is typically used to describe solids featuring fluid or frictional dissipation. \dd{The Kelvin-Voigt model} is governed by the following equation~\cite{lakesvisco}, where $\sigma$ is stress, $E$ is the Young's modulus, $\eta$ is the dashpot stiffness, $\epsilon$ is  strain and $t$ is time:

\begin{equation}
    \sigma=E\epsilon + \eta \frac{\partial \epsilon}{\partial t}.
\end{equation}

The Generalised Maxwell model (Fig. \ref{fig:SI1}B) consists of a spring, with multiple spring-dashpots in parallel and is typically used to describe material dissipation. It is also referred to as the Maxwell-Wiechert model~\cite{Christensenvisco}. The spring-dashpots are each known as Maxwell elements, each with a time scale $\tau_i=\eta_i/E_i$. When only one Maxwell elements is used, the Generalised Maxwell model reduces to the Standard Linear Solid~\cite{lakesvisco}.

When a material based on the Generalised Maxwell model is subjected to a stress relaxation test, the time response of the Young's modulus\dd{, $E(t)$,} is described by:
 \begin{equation}
     E(t) := E_{0} \left(1- \sum_{n=1}^{N} \beta_n \left(1-e^{\frac{-t}{\tau_n}}\right)\right),
     \label{eq:Prony}
 \end{equation}
 with $E_{0}$ the peak Young's modulus under instantaneous load, $\beta_n$ the dimensionless relaxation strength, $\tau_n$ the timescale of the individual Maxwell elements~\cite{Christensenvisco}, $N$ the number of Maxwell elements considered and $E_e = E_{0} \left(1- \sum_{n=1}^{N} \beta_n\right)$ the fully relaxed Young's modulus. Eq. \ref{eq:Prony} is also known as a Prony series. Whether the response at short time scales can be captured largely depends on the loading rate of the test: the higher the loading rate, the shorter the time scales that can be captured.

\bigbreak
\section*{Viscoelastic Properties of Agilus30}
To determine the visco-hyperelastic properties of Agilus30, we performed stress relaxation experiments, which have largely been recorded previously in~\cite{dykstra2019viscoelastic}.  

 An Agilus 30 dogbone section with a length $L$ = 50 mm, depth $d$ = 5 mm and width $w$ = 10 mm has been 3D printed using a Stratasys Objet500 Connex3 printer. The sample is stretched quickly at a strain rate of $\dot{\epsilon} = 0.4\,s^{-1}$, using a uniaxial testing device (Instron 5943) to a strain $\epsilon = 20\%$, after which the force is allowed to relax for one hour. The data is measured with a frequency of 1000 Hz with $t=0 \,s $ defined at the point of highest load.
 
 The parameters of Eqn. (\ref{eq:Prony}) are then fitted using a least-squares fit to the test data, which was interpolated on a logarithmic time scale from 0.01 s to 1 hour. The relaxation test results are presented in Fig. \ref{fig:SI1}C, including fits with $N$=1, $N$ = 3 and $N$ = 5, with corresponding material properties in Tab. \ref{tab:relax}.  
 
\bigbreak

\section*{Switchable Bi-beam}
Here, we provide an example of a model of the strain-rate dependent bi-beams described in Fig. \ref{fig:Figure2}BC~\cite{janbaz2020strain}. The code is available on Zenodo~\cite{Zenodo2022}.

\begin{figure}[h!]
\centerline{\psfig{figure=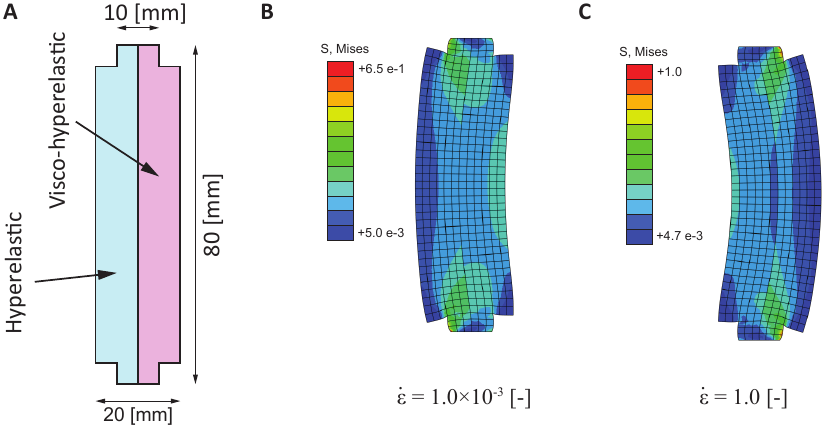,width=.99\columnwidth}}
\caption{\textbf{Bi-beam model.} (A) Bi-beam geometry. (B) Buckling to the left at low strain rates. (C) Buckling to the right at high strain rates. }
\label{fig:ss}
\end{figure}

Bi-beams are made by attaching two similar flexible beams made from two different materials. The left beam  (Fig. \ref{fig:ss}A) is made of a stiff elastic material. The right beam is highly viscoelastic and softer than the elastic one in the relaxed state but stiffer at high strain rates. Such a bi-beam will predictably buckle to either the left (Fig. \ref{fig:ss}B) or the right (Fig. \ref{fig:ss}B) side, depending on the rate of the applied strain. 

\bigbreak
Nonlinear buckling analysis was performed using \dd{the} non-linear visco solver \dd{in Abaqus} (Abaqus 2021. Simulia, USA)~\cite{janbaz2020strain}\dd{: an implicit dynamic solver (see Section \ref{sec:modelling}), which ignores inertial effects}. In order to discretise the geometry of the bi-beam, we used 8-node biquadratic hybrid elements (CPE8H) while 12 elements were seeded through the width of the bi-beam. The bi-beam is clamped at its both ends while the top clamped nodes are moved longitudinally in our simulations. A linear elastic material model (\(E = 1.5\) MPa) was used to define the elastic properties of the left beam and the viscoelastic properties of the right beam is according to the five term Prony series materials parameters that is proposed for Agilus in Table \ref{tab:relax}. The bi-beam in our model has a length and width of 80 mm and 20 mm respectively.

The code shared~\cite{Zenodo2022} can be used to model buckling of bi-beam (to the left) at a strain rate \(\dot\varepsilon = 1\times10^{-3} \) 1/s (Fig. \ref{fig:ss}B) and (to the right) at a strain rate  \(\dot\varepsilon = 1\) 1/s (Fig. \ref{fig:ss}C). It can also serve as a general example on how to model viscoelastic metamaterials.

\bigbreak

\section*{Data and Code Availability}
The data and codes that support the figures within this paper are publicly available on a Zenodo repository~\cite{Zenodo2022}.

\bigbreak
\section*{Acknowledgements}
We acknowledge funding from the European Research Council under grant agreement 852587 and the Netherlands Organi\dd{s}ation for Scientific Research under grant agreement NWO TTW 17883.
\bigbreak

\bibliographystyle{Paper}
\bibliography{Paper}

\end{document}